\documentclass{PoS}
\pdfoutput=1
\usepackage{subfig}
\usepackage[pdftex]{graphicx}
\newcommand{\MSb}{\overline{\mathrm{MS}}}

\title{
\vspace{-2cm}
{\small \normalfont \hfill DESY 11-212\\
  \hfill HU-EP-11/53\\
  \hfill SFB/CPP-11-63\\}
\vspace{1cm}
 Topological susceptibility and chiral condensate with
  $N_f=2+1+1$ dynamical flavors of maximally twisted mass
  fermions. \\
}

\ShortTitle{$\chi_{top}$ and $\Sigma$ for   $N_f=2+1+1$ dynamical flavors of twisted mass fermions.}

\author{K. Cichy $\,\,\,^{a,b}$ , V.
Drach$\,\,^a$, \speaker{E.
    Garcia-Ramos} \thanks{Email: elena.garcia.ramos@desy.de} $\,\,\,^{a,c}$ , K.
Jansen$\,\,^a$\\
\llap{$^a$} NIC, DESY, Platanenallee 6, D-15738 Zeuthen, Germany\\
\llap{$^b$} Adam Mickiewicz University, Faculty of Physics , Umultowska 85,
61-614 Pozna\'n, Poland\\
\llap{$^c$}  Humboldt Universit\"at zu Berlin,  Newtonstrasse 15 12489 Berlin,
Germany\\

}

\abstract{We study the `spectral projector' method for the computation of the 
chiral condensate and the topological susceptibility, using $N_f=2+1+1$ 
dynamical flavors of maximally
twisted mass Wilson fermions. In particular, we  perform a study of the 
quark mass dependence of the chiral condensate $\Sigma$ and
topological susceptibility $\chi_{top}$ in the range 
$270\;\mathrm{MeV} < m_{\pi} < 500\;\mathrm{MeV}$ and compare our data with 
analytical predictions.
In addition, we compute $\chi_{top}$ in the quenched approximation where we 
match the lattice spacing to the $N_f=2+1+1$ dynamical simulations. Using the Kaon, 
$\eta$ and
$\eta'$ meson masses computed on the $N_f=2+1+1$ ensembles, 
we then perform a preliminary test of the Witten-Veneziano relation.}

\FullConference{XXIX International Symposium on Lattice Field Theory
  \\
                 July 10 - 16 2011\\
                 Squaw Valley, Lake Tahoe, California}

\begin{document}

\section{Introduction}

The Banks-Casher relation \cite{BanksCasher} connects the chiral
condensate $\Sigma$ with the density of eigenmodes at the origin 
of the spectrum and thus with the infrared properties of the Dirac operator. 
The chiral condensate is obtained from the eigenvalue density $\rho(\lambda,m)$ 
in a triple limit, 
sending the volume $V$ to infinity and the quark mass $m$ as well as  
the eigenvalues $\lambda$ to zero (in this order),

\begin{equation}
 \frac{\Sigma}{\pi}=\lim_{\lambda\rightarrow0}\lim_{m\rightarrow0}\lim_{
V\rightarrow\infty}\rho(\lambda,m).
\label{eq:bc}
\end{equation}

Only when the eigenvalue density $\rho(\lambda,m)$ 
is non-zero at 
the origin, the chiral condensate does not vanish  
and hence  
the infrared properties of the Dirac operator are directly related to the 
mechanism of chiral symmetry breaking. 

One way to express $\rho(\lambda,m)$ is through the 
mode number $\nu(M,m)$, which is defined as the number of eigenmodes
$\lambda$ of the considered Dirac operator 
squared below some cut-off mass $M$,

\begin{equation}
  \label{nu}
  \nu(M,m)=V\int^{\Lambda}_{-\Lambda}d\lambda\rho(\lambda,m),
  \qquad\Lambda=\sqrt{M^2-m^2} .
\end{equation}

The above considerations can --in principle-- be taken over directly to the lattice  
as a way to compute the chiral condensate non-perturbatively. 
However, counting the eigenmodes below the cut-off $M$ with $M\approx O(100){\rm MeV}$ 
and taking the continuum limit more and more modes have to 
be taken into account for a fixed physical value of $M$. 
In fact, 
a direct counting of the low-lying eigenmodes is expected
to show an $O(V^2)$ scaling behaviour and becomes prohibitively 
computer time expensive when the continuum limit is taken.   

Recently, however, a new method \cite{Giusti} to compute the mode number
was developed, the so-called 
spectral projector
method. The important advantage of this method is that it is 
computationally much faster than counting eigenmodes directly and scales
only 
with the volume $V$.
In addition, the concept of spectral projectors can be extended 
to evaluations of other quantities   
such as the topological susceptibility
$\chi_{\rm top}$ or the ratio of the pseudoscalar and scalar 
renormalization constants $\frac{Z_P}{Z_S}$, as explained in
refs.~\cite{Giusti,Palombi}. 

It goes beyond the scope of this proceedings contribution 
to detail the spectral projector method and we have to refer
to refs.~\cite{Giusti,Palombi} for a description of this method.  
The aim of this contribution is rather to see, how the spectral projector
method works for computing the chiral condensate and the topological
susceptibility in the case of the here used maximally twisted mass 
fermions formulation of lattice QCD \cite{Frezzotti:2003ni}. 
In particular, in this work we 
are interested in the quark mass dependence of the chiral condensate 
and the topological susceptibility and 
we will work at only one value of the
lattice spacing of $a\approx 0.0782$ fm. 
All results are shown for a setup employing a mass-degenerate light 
quark doublet and a strange and a charm quark close to their 
physical values, a situation we refer to as $N_f=2+1+1$, see 
refs.~\cite{Chiarappa:2006ae,Baron:2010bv,Baron:2010th}
for simulation and analysis details.  
Employing several values of the quark mass will allow us to 
confront our data with predictions of chiral perturbation theory 
and to extract values for the chiral condensate in the chiral limit. 
We will also perform a first test of the Witten-Veneziano formula 
\cite{Witten:1979vv,Veneziano:1979ec}  
in this proceedings contribution by computing Kaon, $\eta$ and $\eta'$  
masses on our dynamical $N_f=2+1+1$ configurations and 
the topological susceptibility in the infinite quark mass limit  
(quenched approximation) matched, however, to the physical 
situation of the unquenched simulations. 

\section{Evaluation of the chiral condensate and topological susceptibility}

When computing the chiral condensate from spectral projectors, there 
are two important ingredients. The first is a technical aspect:
the spectral projectors are calculated from a stochastic estimate of 
the ``inverse'' of some suitable function of the lattice Dirac operator
employed, see ref.~\cite{Giusti}. Therefore, it needs to be investigated
what is a sufficient number of stochastic noise vectors employed 
and how the stopping criterion for obtaining the solution of a Dirac 
equation needed to construct the spectral projector.  
The second aspect is more physical and originates from the fact that 
the chiral condensate is computed from the slope of the mode number as a function 
of the cut-off $M$. Before coming to our results, let us therefore briefly discuss 
the tests we have made for both issues. In the following, we will use 
$M^{*}$ as the cut-off for the mode number counting. $M^{*}$ is of a very similar 
size as $M$ and plays the role of an adjustable parameter to optimize
the simulations. 
  
As a very first test, we performed a comparison between the explicitly 
computed mode number and the values from the spectral projectors. 
In this test we found a perfect agreement demonstrating that our implementation 
of the spectral projectors in the tmLQCD package \cite{Jansen:2009xp}
is correct.

\begin{figure}[h]
\begin{center}
\includegraphics[width=0.35\textwidth]{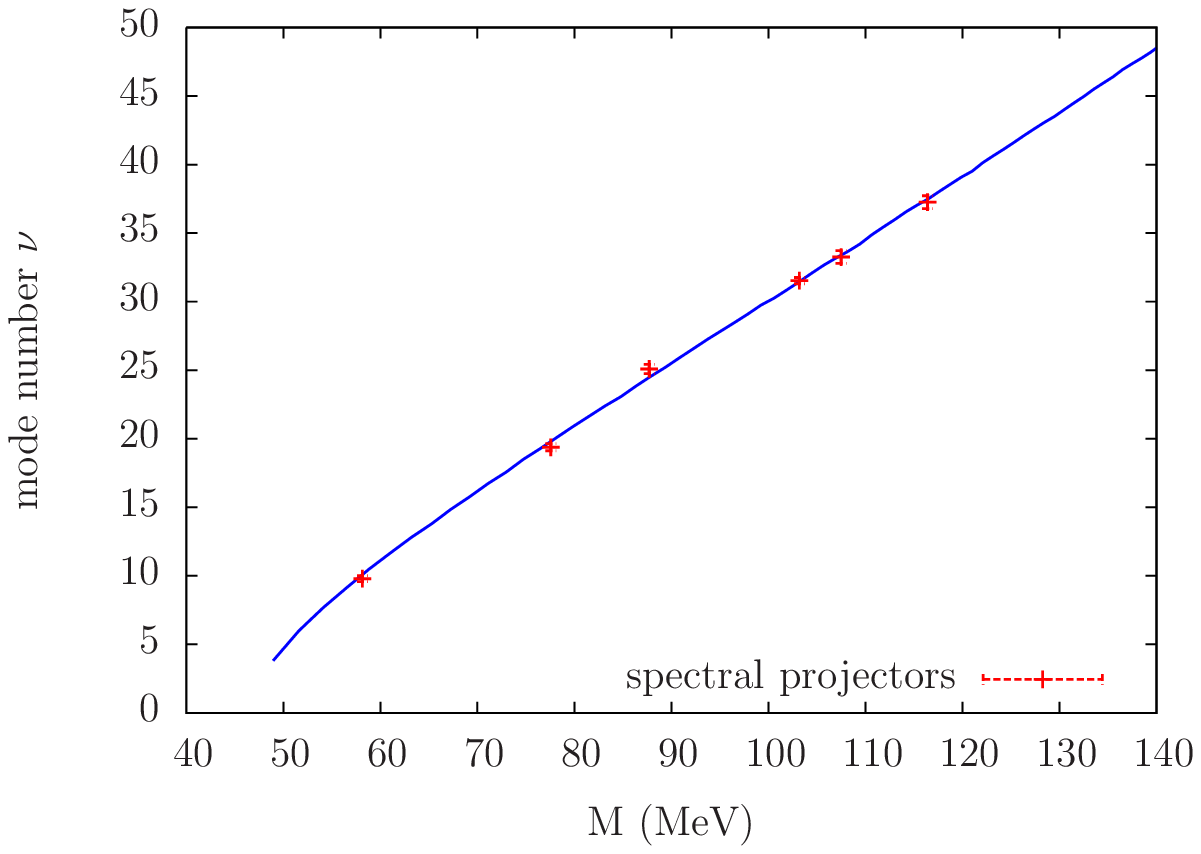} 
\includegraphics[width=0.35\textwidth]{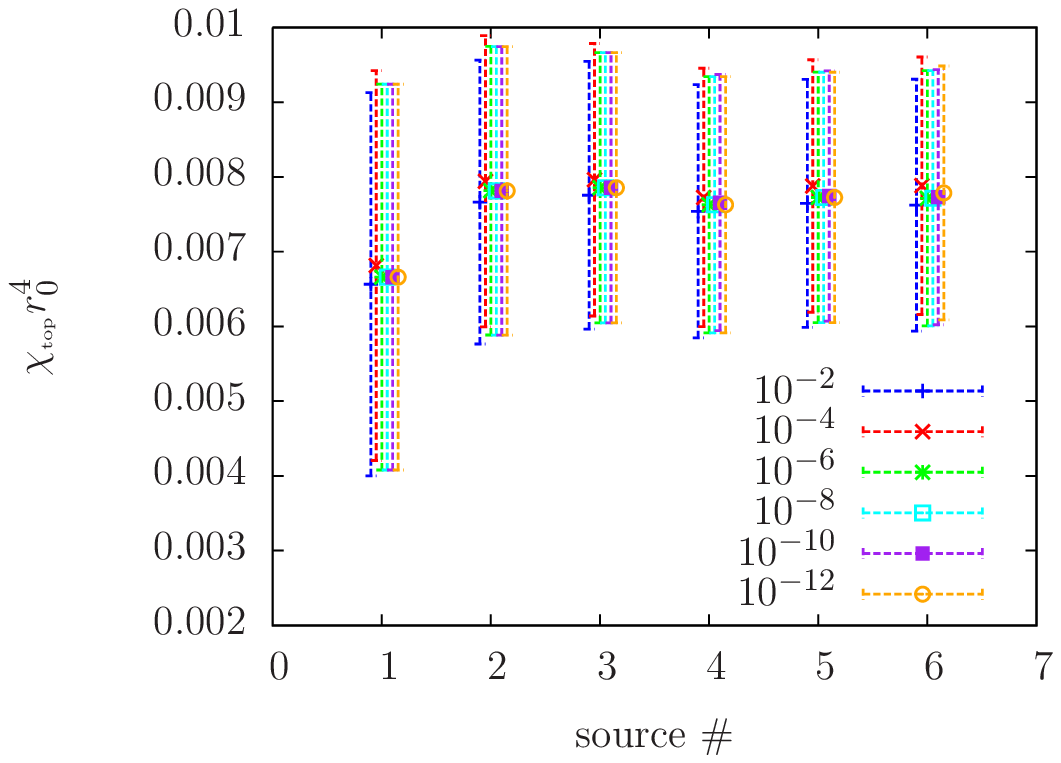}
\caption{\label{test} (left) The mode number as a function of $M$ -- from
explicit computation of eigenmodes (line) and from spectral projectors (points).
(right) The influence of the number of stochastic sources and relative precision
of solving the Dirac equation.} 
\end{center}
\end{figure}

Looking at the mode number 
itself as function of $M$, see fig.~\ref{test} (left),  
we indeed can identify a linear behaviour of the mode number which 
will allow us in the following to extract the chiral condensate. 
We also show in fig.~\ref{test} (right) our results for 
changing the number of stochastic sources and the (relative) stopping 
criterion. As a conclusion from this study we found that with already 
6 stochastic sources the corresponding error saturates, nevertheless
we used 8 sources to remain safe. In addition, we found that 
the stopping criterion can be chosen rather loosely and even 
a choice of $10^{-2}$ gave completely consistent results. Nevertheless, 
for our work we decided to choose a stopping criterion of 
$10^{-6}$ to be on the safe side.  

\vspace{-1cm}

\subsection{Results}

After the tests described in the previous section, we proceeded to compute 
the mode number from spectral projectors. 
As said above, for this work we used only one value of the lattice spacing
of $a \approx 0.0782 {\rm fm}$ corresponding to $\beta=1.95$, as determined by
ETMC \cite{Baron:2010bv}.  
While working here only at one value of the lattice spacing, we have 
computed the mode number at several values of the quark mass, see tab.~\ref{setup}, 
where we give the bare light twisted mass parameter in lattice  
units, as well as the pion mass. 
Note that for all the parameters shown in tab.~\ref{setup} the theory 
was tuned to maximal twist. 
\begin{table}
\begin{minipage}{0.4\linewidth}
\begin{tabular}{cccc}
    \hline
      $\beta$  & lattice & $a\mu_l$ &$ M_{\pi}\; $(MeV)\\
      \hline
      1.95 & $32^3\times64$ & 0.0025 &  270\\
      1.95 & $32^3\times64$ & 0.0035  &320\\
      1.95 & $32^3\times64$ & 0.0055 &390\\
      1.95 & $32^3\times64$ & 0.0075 & 455\\
     1.95 & $24^3\times48$ & 0.0085 &490\\
     \hline
   \end{tabular}
 \end{minipage}
\hspace{0.7cm}
\begin{minipage}{0.55\linewidth}
\vspace*{-1cm}
    \caption{\label{setup} Parameters of ensembles used to
computed the mode
number and the topological susceptibility. We give the light bare 
      twisted mass parameter $a\mu_l$ in lattice  
      units as well as the approximate pion masses. The value 
      of $\beta=1.95$ yields a  
      a lattice spacing of $a\approx0.0782$fm, see ref.~\cite{Baron:2010bv}. \vspace{-0.9cm}}
\end{minipage}
\end{table}
Our typical statistics for computing the mode number and the
topological susceptibility has been 200 configurations
that were separated by 20 HMC trajectories. 

\subsection{Evaluation of the chiral condensate}

We have computed the chiral condensate $\Sigma$ through a study of
the average number of eigenmodes of $D^{\dagger}D$ with
$\lambda<(M^{\star})^2$, computed using the spectral
projector method.

Once the mode number is computed at several values of $M^{\star}$, 
the chiral condensate can be calculated directly from 
the derivative of the mode number with respect to $M^{\star}$ \cite{Giusti},

\begin{equation}
  \label{ch_cond}
\Sigma_R=\frac{\pi}{2V}\sqrt{1-\left(\frac{\mu_{l,R}}{M_R^{\star}}\right)^2}
\frac{\partial}{\partial  M_R^{\star}} \nu_R(M_R^{\star},\mu_{l,R}).
\end{equation}  

In eq.~(\ref{ch_cond}), $\mu_{l,R}$ and $M_R^{\star}$ denote the renormalized 
twisted mass and cut-off parameters. The renormalization constant $Z_P$
($\MSb$, $\mu=2$ GeV) needed for obtaining these renormalized quantities has
been computed by ETMC in a dedicated four flavour simulation, see
ref.~\cite{Dimopoulos:2011wz}. 
Since the mode number is renormalization group invariant
(i.e. $\nu_R(M_R,\mu_{l,R})=\nu(M,\mu_l)$)
\cite{Giusti}, from eq.~(\ref{ch_cond}) we hence obtain directly 
the renormalized chiral condensate at a scale that is inherited 
from $Z_P$. 

\begin{figure}[h]
  \centering
  \includegraphics[width=0.46\textwidth]{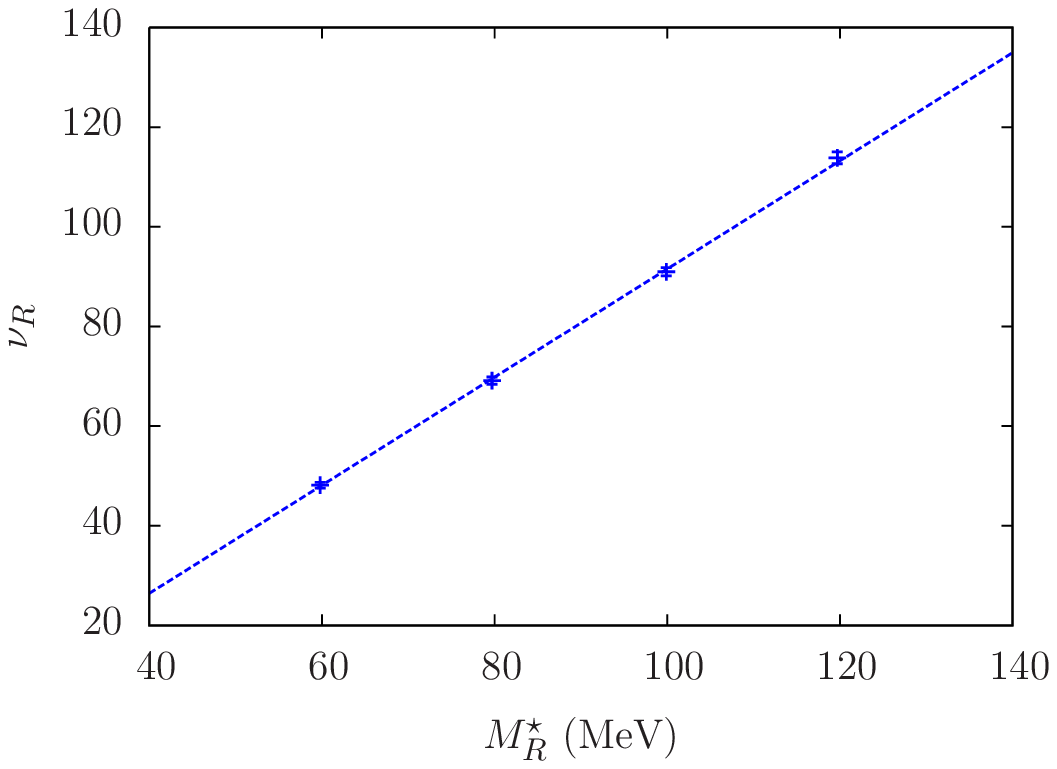}
 \includegraphics[width=0.46\textwidth]{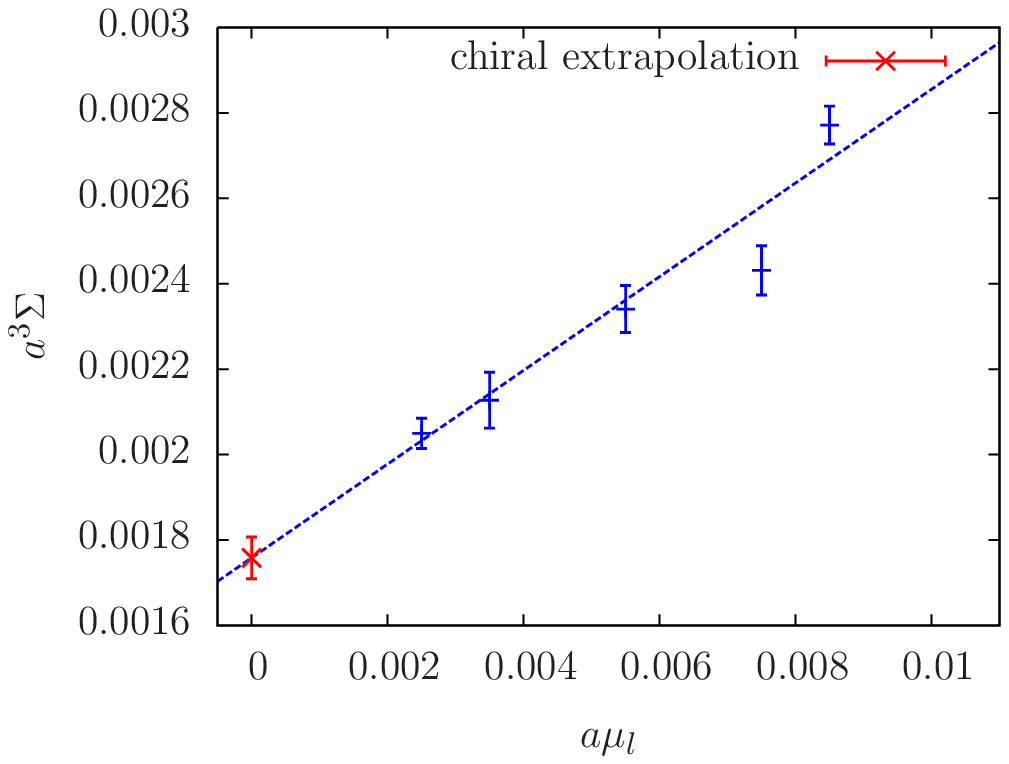}    
\caption{(left) The mode number as a function of $M_R^{\star}$ for
  $\mu_{l,R}\approx 14$ MeV and the corresponding linear fit.
(right) The chiral
condensate $\Sigma_R$ as a function of the renormalized quark mass. The
straight line indicates a linear extrapolation of $\Sigma_R$ to the
chiral limit.}
  \label{nuvsMstar}
\end{figure}

In fig.~\ref{nuvsMstar} (left) we show an example of a behaviour of the
mode number as a function of $M_R^{\star}$, for a renormalized quark
mass of 14MeV. For the four values of $M_R^{\star}$ we have used (and 
similarly as in fig.~\ref{test}), we observe a linear behaviour of
the mode number in the range $60 {\rm MeV} \lesssim M_R^{\star} \lesssim 120 {\rm MeV}$. 
Such linear behaviour of $\nu$ in a range of comparable values of $M_R^{\star}$
was also observed in ref.~\cite{Giusti}. 
From this linear behaviour, which we see at all five quark masses employed, 
we can extract the renormalized chiral condensate $\Sigma_R$ using 
eq.~(\ref{ch_cond}) at a given value of the renormalized $M^\star$. 
In fig.~\ref{nuvsMstar} (right) we show as a result the depence of $\Sigma_R$
on the renormalized quark mass. Extrapolating $\Sigma_R$ linearly to the
chiral limit, we find $\Sigma_R^{1/3}(\MSb,\mu=2\rm GeV)=312(1)(13) MeV$. 
The first error is purely statistical and the
second originates from the 
uncertainty of the renormalization constant $Z_P(\MSb, \mu=2
\rm GeV)=0.462(13)$ \cite{David}.

\subsection{Topological Susceptibility}

As another quantity accessible to the method of the spectral 
projectors, we have computed 
the topological susceptibility $\chi_{\rm top}$ following ref.~\cite{Palombi}. 
In this reference it was demonstrated that besides the bare 
$\chi_{\rm top}$ also the renormalized one can be obtained 
by solely using observables defined through spectral projectors. 
Since we know the necessary renormalization factor, the ratio 
of the scalar to the pseudoscalar renormalization 
constants $\frac{Z_S^2}{Z_P^2}$ available to us from 
the $N_f=4$ simulations of ETMC \cite{David}, we decided to only compute 
the bare value of $\chi_{\rm top}$ from spectral projectors and to perform the renormalization 
using the results from ref.~\cite{David}, i.e. $\frac{Z_P}{Z_S}=0.685$. 

In this way, we have computed $\chi_{\rm top}$ at five values of the 
bare quark mass, listed in tab.~\ref{setup}. 
Before showing our results, we remark that we have performed a test 
of the dependence of $\chi_{\rm top}$ as a function of the cut-off parameter
$M^{\star}_R$. We found that in the range $ 90 {\rm MeV} \lesssim M_R^{\star} \lesssim 130 {\rm MeV}$   
$\chi_{\rm top}$ is constant as a function of $M^{\star}_R$ and we therefore decided
to use a value of $M^{\star}_R\approx 100 {\rm MeV}$ for the computation of all subsequent
values of $\chi_{\rm top}$.

\begin{figure}[h]
\begin{minipage}[b]{0.48\linewidth}
  \centering
 \includegraphics[width=1\textwidth]{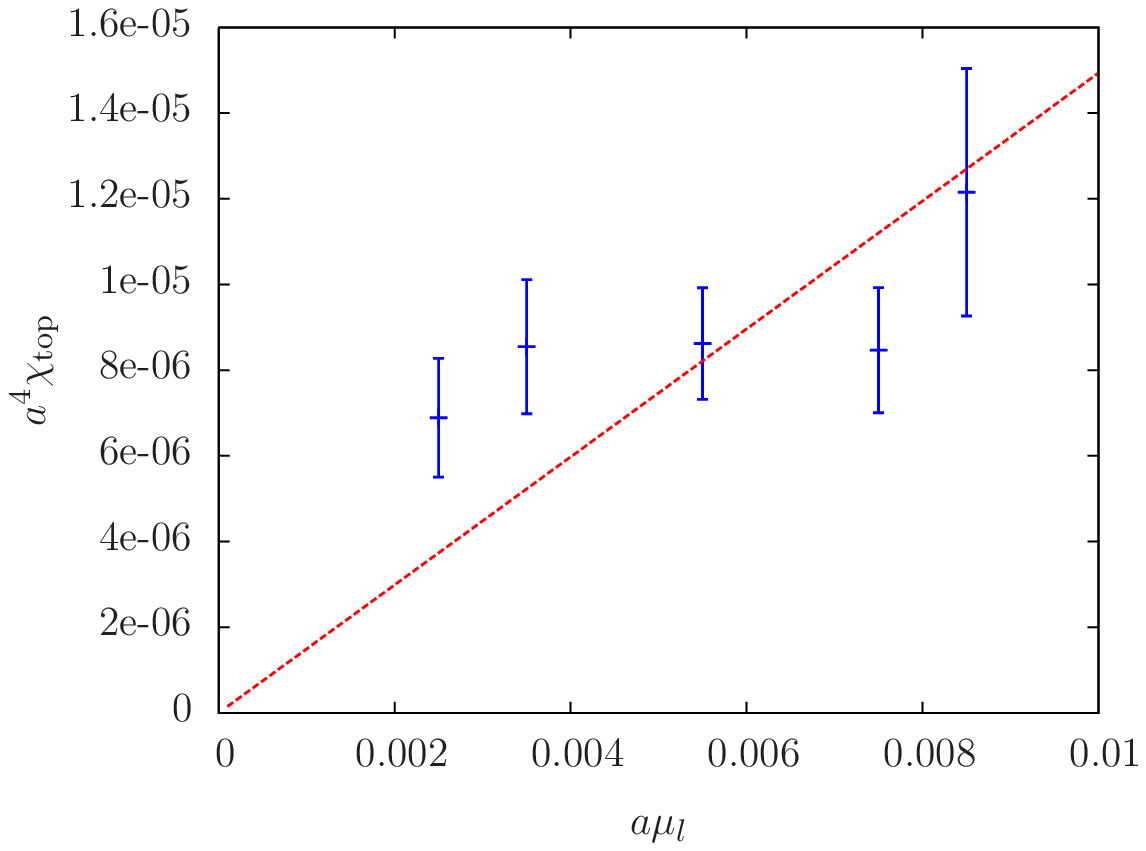}
  \caption{The chiral behaviour of the topological susceptibility with 5
values of the quark mass. The linear fit represents the tree-level formula of
$\chi$PT, $\chi_{top}=\mu_l\Sigma/2$, and yields a value of the chiral condensate.}
\label{ts_ch_limit}
\end{minipage}
\hspace{0.3cm}
\begin{minipage}[b]{0.48\linewidth}
  \centering
 \includegraphics[width=1\textwidth]{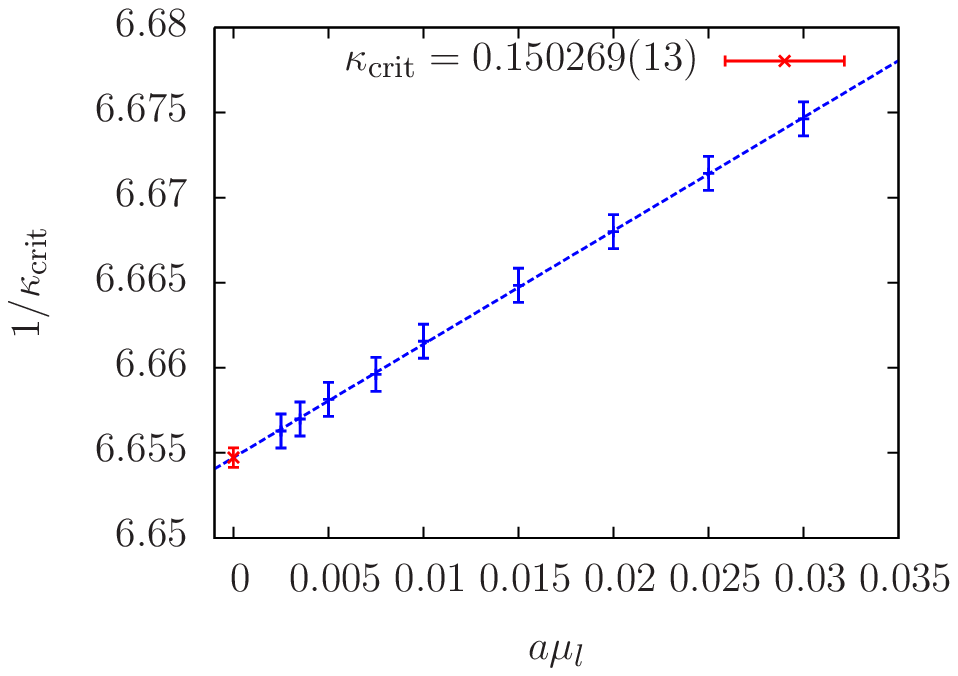}
  \caption{The critical hopping parameter $\kappa_{\rm crit}$
for the quenched
ensemble at $\beta=2.67$ as a function of $\mu_l$. The linear fit yields the
chiral limit value of $\kappa_{\rm crit}$ which is used to realize maximal
twist.}
\label{kappa_c}
\end{minipage}
\end{figure}

Fig.~\ref{ts_ch_limit} shows the chiral behaviour of the topological
susceptibility. We have fitted our results with the tree-level formula of
chiral perturbation theory, $\chi_{top}=\mu_l\Sigma/2$. The fit yields
$\Sigma_R^{1/3}=282(5)(13)$ MeV, where again the systematic error is dominated by
the uncertainty in the renormalization constants.

\section{Witten-Veneziano formula}

As said above, with spectral projectors we have a method at hand 
that allows for a rather cheap computation 
of the topological susceptibility, which is, in addition, well defined, 
i.e. it does not suffer from short distance singularities. 
Moreover, twisted mass fermions are advantageous 
for computing disconnected (singlet or OZI) quantities as, e.g., 
the masses of the $\eta$ and $\eta'$ mesons (see the contribution 
of V.~Drach to this conference and refs.~\cite{Jansen:2008wv,Boucaud:2008xu}). 

Thus, it is very tempting to attempt a non-perturbative test 
of the Witten \cite{Witten:1979vv} -- Veneziano \cite{Veneziano:1979ec}
formula, which provides an elegant 
explanation for the origin
of the unexpectedly large mass of the $\eta^\prime$ meson. 
The Witten-Veneziano (WV) formula relates the masses of the Kaon, $\eta$ and 
$\eta'$ mesons to the topological susceptibility $\chi_{\infty}$ where
the $\infty$ index reminds us that the topological susceptibility 
needs to be computed at infinite quark mass,  i.e. in the pure gauge
theory. 
The formula then reads,
\vspace{-0.3cm}
\begin{equation}
  \label{w-vformula}
  \frac{f_{\pi}^2}{4N_f}\left(m_{\eta}^2+m_{\eta^{\prime}}^2-2m_K^2\right)=\chi_{\infty},
\vspace{-0.3cm}
\end{equation}
where $f_\pi$ is the pion decay constant. Eq.~(\ref{w-vformula}) cannot be
derived in a rigorous way, but some (mild) assumptions are required. It is
therefore most worthwhile to test the formula by direct and non-perturbative
lattice simulations, since the WV formula is one of the most fundamental
relations in QCD and clearly points out the importance of topology.

\subsection{Strategy to compute $\chi_{\infty}$}

In order to test the WV formula (\ref{w-vformula}),
the topological susceptibility $\chi_{\infty}$ needs to be 
computed in the pure gauge theory, but at a matched physical situation. 
To fulfil this condition, we performed a scan in $\beta$ using the same 
(Iwasaki) gauge action as used in \cite{Baron:2010bv} and searched for the
value of $\beta$ that gives the same value of the force parameter $r_0$
as the (chirally extrapolated) one of our dynamical ensembles at $\beta=1.95$.
As a result, we obtained $\beta=2.67$. 
The meson masses that are needed in the WV formula were evaluated for 
a bare twisted mass parameter of $\mu_l=0.0055$. 

Since the topological susceptibility from spectral projectors 
is a fermionic quantity and we want to use maximally twisted 
mass fermions, a first step is the tuning to maximal twist.
This amounts to tuning the bare Wilson quark mass, or equivalently
the hopping parameter $\kappa$, to its critical value, $m_{\rm crit}$ ,
leading to $\kappa_{\rm crit}=1/(8+2m_{\rm crit})$. 
Following the strategy introduced in ref.~\cite{kappa}, we have computed
$\kappa_{\rm crit}$ at different values of the twisted mass $\mu_l$ 
by demanding that 
the PCAC quark mass vanishes. In the end, we have 
performed a chiral extrapolation letting $\mu_l$
approach zero. Our critical value 
of $\kappa_{\rm crit}$ is then the one in the chiral 
limit. The results of this procedure are shown in
fig.~\ref{kappa_c}, which shows that a linear extrapolation to zero 
twisted mass parameter is justified. 

Since in the pure gauge theory and with the Iwasaki gauge action the renormalization 
constants $Z_S$ and $Z_P$ are not available to us, we decided to follow
in this case refs.~\cite{Giusti,Palombi} and compute 
the renormalized $\chi_{\infty}$ solely from suitable 
expectation values employing spectral projectors.

In this way, we finally obtained a value of $\chi_{\infty}$ 
which is listed in tab.~\ref{W-V} together with our results of the 
meson masses and $f_\pi$ relevant for the WV formula. 
Putting everything together and multiplying both sides 
of eq.~(\ref{w-vformula}) with $r_0^4$ to make it dimensionless,
 we find for the left hand side of the 
WV formula 0.036(8) and for the right hand side 0.053(18). 
Although within the errors the WV formula is fulfilled, 
our present data clearly do not allow us to perform a stringent test.  

\begin{table}[h!]
\hspace{-0.1cm}
 \begin{minipage}{0.6\linewidth}
    \begin{tabular}{|c|c|c|c|c|}
    \hline
      
      \hline
      $am_{\eta}$  & $am_{\eta^{\prime} }$& $am_K $&       
      $af_{\pi}$ & $a^4\chi_{\infty} $ \\
      \hline
     0.230(10) & 0.384(24) & 0.2280(4) & 0.0656(2) &0.000050(17)\\
     \hline   
    \multicolumn{4}{|c|}{ $r_0^4\frac{f_{\pi}^2}{4N_f}\left(m_{\eta}^2+m_{\eta^{\prime}}^2-2m_K^2\right)=0.036(8)$}&
$ r_0^4\chi_\infty=0.053(18)$\\
      \hline
    \end{tabular}
  \end{minipage}
\hspace{2cm}
  \begin{minipage}{0.25\linewidth}
    \caption{\label{W-V} Results of the meson masses and $\chi_{\infty}$
 in pure gauge theory.}
  \end{minipage}
\end{table}

\vspace*{-1.2cm}
\section{Conclusion} 
\vspace{-0.2cm}
In this proceedings contribution we have explored the potential
of the newly introduced spectral projectors method when applied 
to maximally twisted mass fermions in the $N_f=2+1+1$ setup. 
Our analysis has used only one value of the lattice spacing, but several 
quark masses which allowed us to compute the chiral condensate 
in the chiral limit, both from the quark mass dependence of the 
condensate itself and the topological susceptibility. In addition, we can compare these
values to the ones of ref.~\cite{Baron:2010bv}, where the chiral condensate 
has been extracted from chiral perturbation theory fits 
to the pion mass and decay constant. In tab.~\ref{results} 
we show these different results. The agreement
between the extraction of $\Sigma$ using very different methods
is reassuring.
\begin{table}[h!]
  \begin{minipage}{0.45\linewidth}
    \begin{tabular}{|c|c|c|}
     \hline
      \multicolumn{3}{|c|}{$\Sigma^{1/3}_R(\mu=2GeV) [MeV]$}\\
      \hline
     spectral proj &  chiral fits& $\chi_{top} $ \\
      \hline
      312(1)(13) & 276(2)(11)& 282(5)(13)\\
      \hline
  \end{tabular}
 \end{minipage}
\hspace{0.5cm}
  \begin{minipage}{0.5\linewidth}
 \caption{\label{results} Results for the chiral condensate obtained
    with 3 different methods.}   
  \end{minipage}
 \end{table}

\vspace{-0.2cm}

As another step, we have performed a first test of the Witten-Veneziano formula.
Unfortunately, 
at the moment our accuracy does not allow for a stringent test of this 
fundamental relation between meson masses and the (quenched) topological 
susceptibility. However, the fact that we found reasonable errors already is 
quite promising that in the future a more precise test can be performed. 
Clearly, in this work a number of systematic effects could no be considered yet. 
In particular, it will be interesting to understand the effects of the 
lattice spacings. 

As a last remark, we want to mention that the spectral projector expectation values 
used to compute the topological susceptibility show a high sensitivity 
of autocorrelations stemming from topology. Since these expectation values
are rather cheap to compute, it can thus be envisaged that 
such quantities can be used to scrutinize simulations 
of lattice QCD for possible large autocorrelation times. 
\vspace{-0.2cm}

\section*{Acknowledgments}
\vspace{-0.2cm}
E. Garcia-Ramos was supported by the Deutsche Forschungsgemeinschaft in
the SFB/TR 09. K. Cichy was supported by the Kolumb
fellowship, financed by the Foundation for Polish Science.
The computer time for this project was made available to us by the Juelich
Supercomputing Center, the PC cluster in Zeuthen, LRZ in
Munich, Poznan Supercomputing and Networking Center. We thank these computer
centers and their staff for all technical advice and help. We also
thank all members of ETMC for useful discussions, in particular G. Herdoiza.


\end{document}